\documentclass{ws-jin}
\usepackage{graphicx}
\usepackage[cp1251]{inputenc}
\inputencoding{cp1251}
\usepackage{url}
\usepackage{wsnatbib}
\let\cite=\citep
\usepackage{color}
\definecolor{darkgreen}{rgb}{0,.5,0}
\usepackage[unicode,colorlinks,filecolor=blue,citecolor=darkgreen]{hyperref}
\begin{document}
\def\bib{B\kern-.05em{I}\kern-.025em{B}\kern-.08em}
\def\btex{B\kern-.05em{I}\kern-.025em{B}\kern-.08em\TeX}

\markboth{Suleymanov, Gafarov, Khusnutdinov}{Modeling of Interstitial Branching of Axonal Networks}

\catchline{}{}{}{}{}
%%%%%%%%%%%%%%%%%%%%%%%%%%%%%%%%%%%%%%%%%%%%%%%%%%%%%%%%%%%%%%%%%%%%

\title{MODELING OF INTERSTITIAL BRANCHING OF AXONAL NETWORKS}

\author{Y. SULEYMANOV, F. GAFAROV\footnote{Corresponding author. Mail to fgafarov@yandex.ru}}
%\address{Insitute of Information Systems, Kazan Federal University, Kremlevskaya 18, Kazan, 420008, Russia\\
%\email{yaniscom@mail.ru}}

%\author{F. GAFAROV}
\address{Institute of Information Systems, Kazan Federal University, Kremlevskaya 18,
Kazan, 420008, Russia}

\author{N. KHUSNUTDINOV}
\address{Institute of Physics, Kazan Federal University, Kremlevskaya 18, Kazan, 420008, Russia\\
7nail7@gmail.com}

\maketitle

\begin{history}
\received{Day Month Year}
\revised{Day Month Year}
\end{history}

\begin{abstract}
A single axon can generate branches connecting with plenty synaptic targets. Process of branching is very important for making connections in central nervous system. The interstitial branching along primary axon shaft occurs during nervous system development. Growing axon makes pause in its movement and leaves active points behind its terminal. A mathematical model is developed to describe and investigate axonal network branching process. The model under consideration describes axonal network growth in which the concentration of axon guidance molecules manages axon's growth. We model the interstitial branching from axon shaft. Numerical simulations show that in the model framework axonal networks branch similarly to real neural networks in vitro.
\end{abstract}

\keywords{axon; branching model; interstitial; neuron; activity; growth cone}

\section{Introduction}

Processes underlying brain self-organization and operation of the cerebral cortex is still unclear. Different theoretical approaches describing complex spatio-temporal structures and connections were suggested  \cite{Andras:2003:Amfecoisnn,Chauvet:2002:Otmiotntbotsfit,Chauvet:2002:Otmiotntbotsfinsfma,%
Hentschel:1999:Moagabdd,Pelt:2005:DDaPoNNfMtFC,Segev:2000:Gmocbsonn}. These investigations give new understanding of the phenomena observed in neural network formation stage.

Formation of specific nervous system crucially depends on the possibility of creation contacts between sets of neurons. For this reason, the main problem of neurobiology becomes revealing mechanisms for recognition of the target neurons by axons.

Evolution of the neuron's connections is important for correct functionality of nervous system. Nervous system development is accompanied by axons branching out, making contacts with plenty target neurons \cite{Gibson:2011:Droabitvns}. The importance of understanding mechanisms of temporal and spatial  axon's branching when investigating neural networks is yet unknown with precision. 

Axonal branches may be characterized by their morphology, complexity and branch function which they produce.  One shape of branching is arborization. The formation of arborization usually takes place in axon's terminal end in the target region and leads to formation of arbor branches. Another shape of branching is formed by bifurcation of growth cone. As the result two daughter branches appear. They move away from each other as was observed in center sensor projections in spinal cord. Besides these types of branching there is  an interstitial branching which usually is formed at some distance from the axon's terminal and grows toward the target neuron from main axon. This form of branching is produced by sensory collaterals in the spinal cord and by the descending projections from the cortex \cite{Gibson:2011:Droabitvns}.

In the paper we suggest a mathematical model which describes growth of neural network taking into account the axons branching process. The model is based on the experimental biological data. We use interstitial branching of the axons -- the branch appears on the axon's shaft behind growth cone \cite{Bastmeyer:1996:Dotrbiabadca,Dent:1999:RaMoMiAGCaDIB,Dent:2003:Agbgcabccasm,Kalil:2000:Cmugcgaab,%
Soussi-Yanicostas:2002:Aditxfokspabffobon,Szebenyi:1998:Ibdfarotadbtpgcdpb}. The  interstitial branching happens not by  biffurcation but by reorganization of the growth cone because after reorganization the growth cone still continues its movement. After reorganization of growth cone, on axon's shaft appears a place usually called branching point.  Here new branch is formed as a response to external molecular signals. The main novelty of our approach is consideration of branching process and network growth depending on axon guidance molecules (AGM) concentration. The paper is continuation of our previous investigations \cite{Gafarov:2006:Sinnopcnftrcd,Gafarov:2009:Mnctdocc}.

\section{Axon branching processes}

Axon branching, along with its growth and guidance is tightly controlled and requires concordant activity of many processes \cite{Gibson:2011:Droabitvns}. 

Evolving growth cones are directed to the corresponding target neurons. Some cortex neuron pathways emerge by the evolution of the interstitial branching from the axon's body. The interstitial branching appears from the axon shaft sometimes very far behind the main growth cone. The branching process is closely connected with delay of the growth cone. The duration of delay can take  from few hours to few days. In this time the expansion of the growth cone takes place. When the cone starts moving again, the part of the filopodium and lamellipodium which remains on the axon behind its growth cone organizes interstitial branching. In natural conditions cerebral cortical axons innervate their target neurons via interstitial branching from axons shaft \cite{Bastmeyer:1996:Dotrbiabadca,Kalil:2000:Cmugcgaab,Ruthazer:2010:RoibitdovccAtafa}.

One possible mechanism of interstitial branch formation is that axon shaft itself replies for signals taken from the target which are independent from the main growth cone. Another possibility is that the main growth cone recognizes the target and  demarcates specific domains of axon for future branching. Pauses of the cortical neurons growth cones movements take place in regions where axon branches appear later. In the pausing time the growth cone reorganizes and forms new growth cone. The rests of the reorganized cones are left behind on the axon shaft as active filopodium and new branches grow  from these points \cite{Szebenyi:1998:Ibdfarotadbtpgcdpb}. So, the question is not only who recognizes  "branching signal"\/ but also what type of  signal is that. Factors released by target neurons can define the place of axon branching. Factors from local tissues can provide an instructive cue to promote branch formation along the axon. For example, the nerve growth factor (NGF) molecules  may trigger the growth of the axons interstitial branches growing \cite{Gallo:1998:LSoNIACS}. Similar effect was observed with fibroblast growth factor 2 (FGF-2) on pyramidal neuron's axons, but in this case FGF-2  affected region nearby of the growth cone\cite{Szebenyi:2001:Fgfpabocnbimabotpgc}. In this case the axon branching of separate cortical neurons takes place without targets and application  of FGF-2 just increases branching \cite{Kalil:2000:Cmugcgaab}. The guidance molecule Netrin-1 may stimulate the formation of the cortical axon branching  \cite{Dent:2004:NaS3PoICABRbRotC}. The influence  of Netrin-1 on the branch process formation stimulates space-bounded $\textrm{Ca}^{2+}$ processes in axon that take place in the moment of branching formation \cite{Tang:2005:Niabidcnbfcsp}.  Moreover, the local application the Netrin-1 stimulates local $\textrm{Ca}^{2+}$  transition process accompanied by growth of the stimulated branching \cite{Hutchins:2008:DOoAatBIRbLCT}. At last the protein Anosmin commonly known as KAL1 can assist formation of local interstitial branches from projection neurons in the mammalian olfactory bulb \cite{Soussi-Yanicostas:2002:Aditxfokspabffobon}.

As was shown by fluorescently labeled microtubules and high resolution images, the process of branch formation takes place in the growth cone. The growth cone distinguishes the future branching points. In the rests left behind the branches can appear later \cite{Kalil:2000:Cmugcgaab}.

Growth of the interstitial branches is accompanied by reorganization of microtubules (MT) from constrained sets to diffused sets and transformation of long MT to short MT. Long-term reorganizations of MT is accompanied by cancellation of some axonal processes and increasing and stabilisation of others. The MT move inside axonal growth cones developing interstitial branches. During the branch outgrowth from the axons shaft the MT are reorganized in more plastic form, they may go alone out and may be fragmented. Short MT break into developing branches by anterograde and retrograde movements \cite{Dent:1999:RaMoMiAGCaDIB}.

Axons branching process also depends on the neuron's activity which may increase or decrease the branching processes. The investigations made previously with cats show that the blocking of the neural activity by tetrodotoxin significantly decreases the area occupied by axon arbors of thalamocortical fibres \cite{Herrmann21111995}.  The studies of blocking rat's cortical neurons  activity with pharmacological inhibitors reveals the decreasing of the branches number only expansion of arbors maintained the same level \cite{Uesaka:2005:Adocabfamaesuosc}. Therefore, regulation of axon activity can enlarge or diminish  the size of arbors as well as it can change the complexity of structure.

\section{The model of growth and branching}

Direction of growth cones motion is controlled by diffusible chemicals -- axon guidance molecules (AGM) \cite{Goodhill:1997:Diag}. Concentration $c_i = c_i(\textbf{r}_j - \textbf{r}_i,t)$ of AGM  at the point $\textbf{r}_j$ in the moment $t$ depends on the concentration of AGM released by $i$-th neuron at the point $\textbf{r}_i$. Concentration of AGM obeys standard diffusion equation
\begin{equation}
 \frac{\partial c_{i}}{\partial t} - D^2 \triangle_d c_{i} + k c_{i} =
J_i(\textbf{r}_j,t).
\end{equation}
Here $D^2$ -- the diffusion coefficient, $k$ is the degradation coefficient and $\triangle_d$ is the Laplace operator in $d$-dimensional space. Quantity $J_i(\textbf{r}_j,t)$ means some external source. Solution of this equation is well-known for a long time and it has the following form:
\begin{equation}
 c_i(\textbf{r} - \textbf{r}_i,t) = \int G_d (\textbf{r} - \textbf{r}',t) c_i^0(\textbf{r}')d\textbf{r}' + \int_0^t dt'\int G_d (\textbf{r} - \textbf{r}',t-t') J_i(\textbf{r}',t')d\textbf{r}',
\end{equation}
where $c_i^0(\textbf{r})$ -- initial distribution of concentration and $G_d(\textbf{r},t)$ -- the Green function in dimension $d$:
\begin{equation}
 G_d(\textbf{r},t) = \frac{1}{(4\pi t D^2)^{d/2}}
e^{-kt -\frac{\textbf{r}^2}{4tD^2}}.
\end{equation}
At the initial moment $t=0$ there is no AGM, $c_i^0(\textbf{r}) = 0$, and the source
\begin{equation}
 J_i(\textbf{r}_i,t) = a \delta^{(d)} (\textbf{r}_j - \textbf{r}_i) j_i(t),
\end{equation}
is localized on $i$-th neuron. Parameter $a$ describes the amount of AGM released by neuron per second.  The function $j_i(t)$ is perceptivity of $i$-th neuron at the moment $t$ to AGM, that hereafter we will name as activity. We suppose that $j_i(t)\leq 1$. 

Therefore, we arrive with the following expression for concentration
\begin{equation}
 c_i(\textbf{r}_j - \textbf{r}_i,t)= a\int_0^t dt' G_d (\textbf{r}_j -
\textbf{r}_i,t-t') j_i(t').\label{eqconc}
\end{equation}
Because  of the activity of the neurons has no relaxation form but should depend on the surrounded neurons we  adopt the following equation for activity,
\begin{equation}
\tau \frac{dj_i(t)}{dt} = -j_i(t) + f\left(j^{ext}_i(t) + \sum_{k=1,k\not =i}^N
\omega_{ik} j_k(t)\right),\label{eqactiv}
\end{equation}
where $f(x) = x \theta (x)$ and $\theta (x)$ is step function. Here $j^{ext}_i(t)$ -- external source depending on the time $t$, $j_i(t)$ -- activity of $i$-th neuron, $\tau$ -- relaxation time of activity. The $\omega_{ik}$ are   weights which define influence of $k$-th neuron on the $i$-th neuron and may take three value, $-1, 0$ and $1$. If  $\omega_{ik}=1$ then it is excitatory connection and vice versa if $\omega_{ik}=-1$ then it is inhibitory connection, $\omega_{ik}=0$ means no influence. Because neuron can't make connection with itself then $\omega_{ii}=0$. For  $j_i(t) > j^{th}$ the $\omega_{ik}=-1$ and for $j_i(t) < j^{th}$ the $\omega_{ik}=1$. Here $j^{th}$ is a threshold value for activity.

The $i$-th axon growth dynamic is described by vector $\textbf{g}_i$ of the axon's terminal which obeys to differential equation
\begin{equation}
 \frac{d\textbf{g}_i}{dt} = \lambda \theta(j_i - j^{th})\sum_{k=1}^N \nabla
c_k(\textbf{g}_i - \textbf{r}_k,t).\label{eqaxon}
\end{equation}
Axon moves if $j_i > j^{th}$. Parameter $\lambda$ describes axon's sensitivity. At initial moment the coordinates of axon's terminal are equal to the coordinates of its neuron and activity $j_i(t) =0$. Also there are no connections between  $k$-th and $i$-th neurons and all  $\omega_{ik}=0$.

In the framework of model branching process has two stages. At the first stage we find points on the axon's shaft in which the future branch will grow up and we define direction for growth. To find the points we calculate its growth rate, concentration of AGM on its terminal and its length. Branching starts if three conditions are fulfilled:
\begin{itemize}
 \item[(i)] Velocity of its terminal should be smaller than some threshold value $v_g$,
 \item[(ii)] Concentration of AGM at the terminal should be in range between $c_{min}$ and $c_{max}$,
 \item[(iii)] Length $L_b$ of axon's shaft should be greater than threshold value $L_{th}$.
\end{itemize}
\begin{figure}[ht]
\includegraphics[width=10cm]{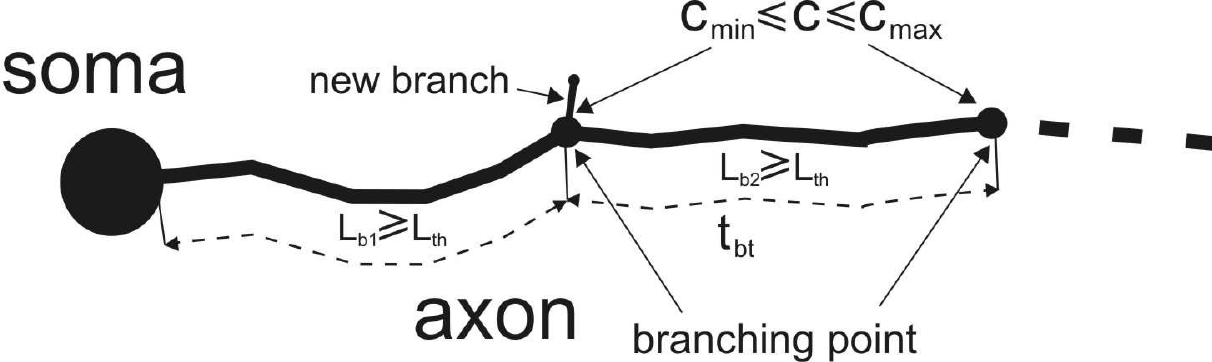}
\caption{Axon with branching points. Here $L_{b1}$ and $L_{b2}$ -- lengths of the first and the second segments of axon between branching points which should exceed threshold value $L_{th}$ in order to branching process may start.  The interval of time $t_{bt}$ means the time after what the mechanism of branch formation starts from branching point. $C$ is concentration axon guidance molecules at the moment of creation of branching point. The $C_{min}$ and $C_{max}$ mean the interval for concentration where formation of branch is possible.}
\label{fig:1}
\end{figure}

\begin{figure}[h]
\includegraphics[width=12cm]{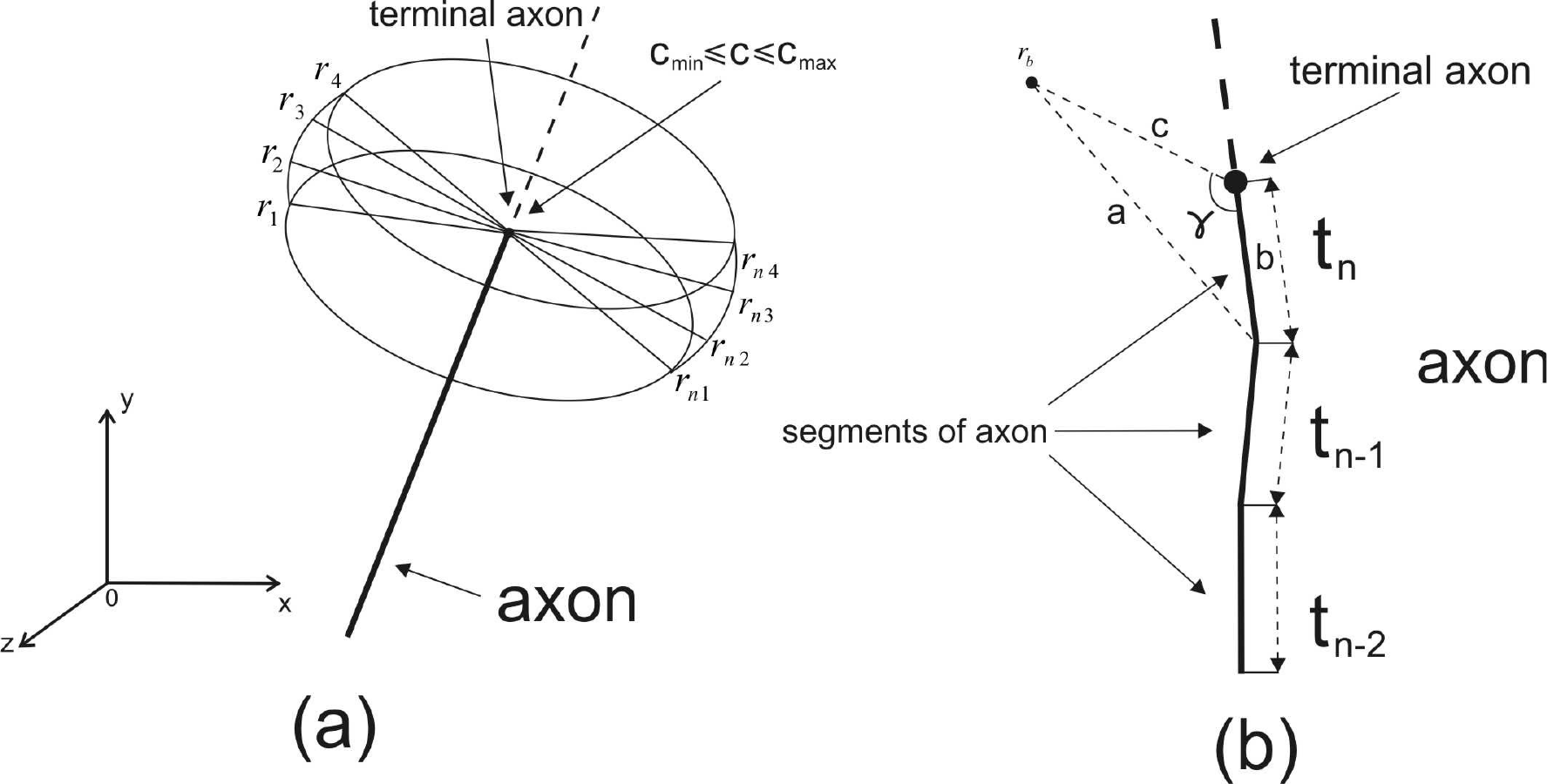}
\caption{The procedure of finding point for new branch in $3$-dimensional space. (a) Calculation the concentration around axon's terminal. (b) Calculation of angle between shaft of axon and the line between terminal of axon and the point of observation. Axon is shown with 3 segments, that were growing during  the time $\Delta t$, the angle $\gamma$ and triangle for calculation angle $\gamma$.}\label{fig:2}
\end{figure}

If these conditions are fulfilled the coordinates of this point are stored. New branch starts to growing after some time after growth cone reorganization \cite{Dent:2004:NaS3PoICABRbRotC,Kalil:2000:Cmugcgaab} so, we launch timer for interval of time $t_{bt}$. New branch does not grow during this time. After this period of time  concentration of AGM is verified at the branching point. If the concentration of AGM belongs interval from  $c_{min}$ up to $c_{max}$ (see Fig. \ref{fig:1}), the second stage starts. The axon's branching process depends on the activity and the greater it is, the greater is the intensity of branching becomes  \cite{Uesaka:2007:Iblsaamotab,Uesaka:2005:Adocabfamaesuosc,Ohnami:2008:Roriacab}. Plenty branching points appear on the axon's shaft during its growth, but only part of them serve as the  basis for branches.  For this reason we consider this process as random process \cite{Koene:2009:NAfftsgolsnnwrnm}. Branching process starts if the following condition is fulfilled:
\begin{equation}
\frac{1}{j^{th}}j \ge P,\label{eqprobability}
\end{equation}
where $P$ -- a random number obtained by generator of the random numbers $0\leq P \leq 1$ and $j$ is the axon's activity. Observations show that the growth cone "touches"\/ the space around itself \cite{Aeschlimann:2000:BMoAP}. In analogy of this we calculate AGM concentration  close to terminal in some points taking into account Eq. (\ref{eqconc}) (see Fig. \ref{fig:2}(a)). Coordinates of these points are calculated with help of equations
\begin{equation}
x_i=x_a+r_b\sin \theta \cos \varphi\,,\
y_i=y_a+r_b\sin \theta \sin \varphi\,,\
z_i=z_a+r_b\cos \theta\,,
\label{xyz}
\end{equation}
where $x_i,y_i$ and $z_i$ -- coordinates points on the sphere, $x_a,y_a$ and $z_a$ -- coordinates of the axon's terminal and the step of angles $\theta$ and $\varphi$ is $5^\circ$. Next we calculate the AGM concentration  at the distance $r_b$ from axon's terminal at point $r_i$ and calculate angle $\gamma$ between shaft of axon and line between point $r_i$ and terminal of axon (see Fig. \ref{fig:2}(b)). If this angle belongs interval from $\gamma_{min}$ to $\gamma_{max}$ we add new branch in the set of branches.  From this set of branches we select single branch with maximum  AGM concentration. Therefore, we obtain new branch of axon.

\section{Numerical simulations}

In the framework of model we neglect the geometry of neuron's soma. We consider it as sphere or circle depending on dimension of problem under consideration. Also we suppose that all neurons are identical with respect their form and behavior, the positions are fixed. We will call these spheres as neurons.  

Let us consider axonal network which consists of the $12$ neurons and then the network composed of $18$ neurons. The parameters used in the paper are shown in the Table \ref{tab:tab-1}.

\begin{table}[h]
\tbl{Parameters used for numerical calculations.}
{\tabcolsep0.3cm\begin{tabular}{@{}p{.69\textwidth}lc@{}}
\Hline\\[-6pt]
Parameter & Value \\[3pt]
\hline\\[-6pt]
Amount of AGM per unit second,  \cite{Goodhill:1997:Diag,Gundersen:1980:Cottrodrntngf} & $a=10^{-5} \frac{nM}s$ \\
AGM diffusion coefficient, \cite{Goodhill:1997:Diag} & $D^2=6\cdot 10^{-7} \frac{cm^2}s$  \\
Relaxation time of activity, \cite{Vogels:2005:Nnd} & $\tau=1 s$  \\
Threshold parameter,  \cite{Segev:2003:Foeacnn} & $j^{th}=0.51$ \\
Growth's speed of an axon for branching & $v_g=5\cdot 10^{-7} \frac{cm}s$ \\
Degradation coefficient & $k=10^{-3} s^{-1}$ \\
Radius of soma & $r=5\cdot 10^{-3} cm$ \\
Minimum concentration of substance for branching & $c_{min}=1\frac{nM}{cm^3}$ \\
Maximum concentration of substance for branching & $c_{max}=30 \frac{nM}{cm^3}$ \\
Coefficient describing axons sensitivity and motility, \cite{Rosoff:2004:Ancastesoatmg} & $\lambda=4\cdot 10^{-6} \frac{cm^5}{nM\cdot s}$
\\[3pt]
\Hline
\end{tabular}}\label{tab:tab-1}
\end{table}

In the first experiment we used network from $12$ neurons with different distances between them. We made experiments $6$ times for different values of the external activity of  neurons $j_{ext} =0.05, 0.1,0.2,0.3,0.4,0.5$. At the beginning the activity of all neurons was the same and equal $j_{ext}$, which models growth of axonal network with blockades such as TTX, APV and DNQX. For each value of the external activity we made the numerical experiment $9$ times, because the branching process has causal origin and connects with neuron activity. We used following parameters: initial length of the branch $r_{b}=1.5\cdot 10^{-3} cm$, angle $\gamma_{min}=90^\circ$ and  $\gamma_{max}=95^\circ$, the time $t_{bt}=25000 s$ and the threshold for minimal length of axon's segment  $L_{th}=5\cdot 10^{-2} cm$.

In framework of our model neurons may have activity from $0$ to $1$. Neurons with activity around zero are inactive neurons. These neurons both have not been stimulated by the external signals and activity of these neurons is depressed by chemicals. Therefore, small values of neuron's activity models the axonal network in the chemical surround such as TTX, APV and DNQX.

\begin{figure}[ht]
\includegraphics[width=6.5cm]{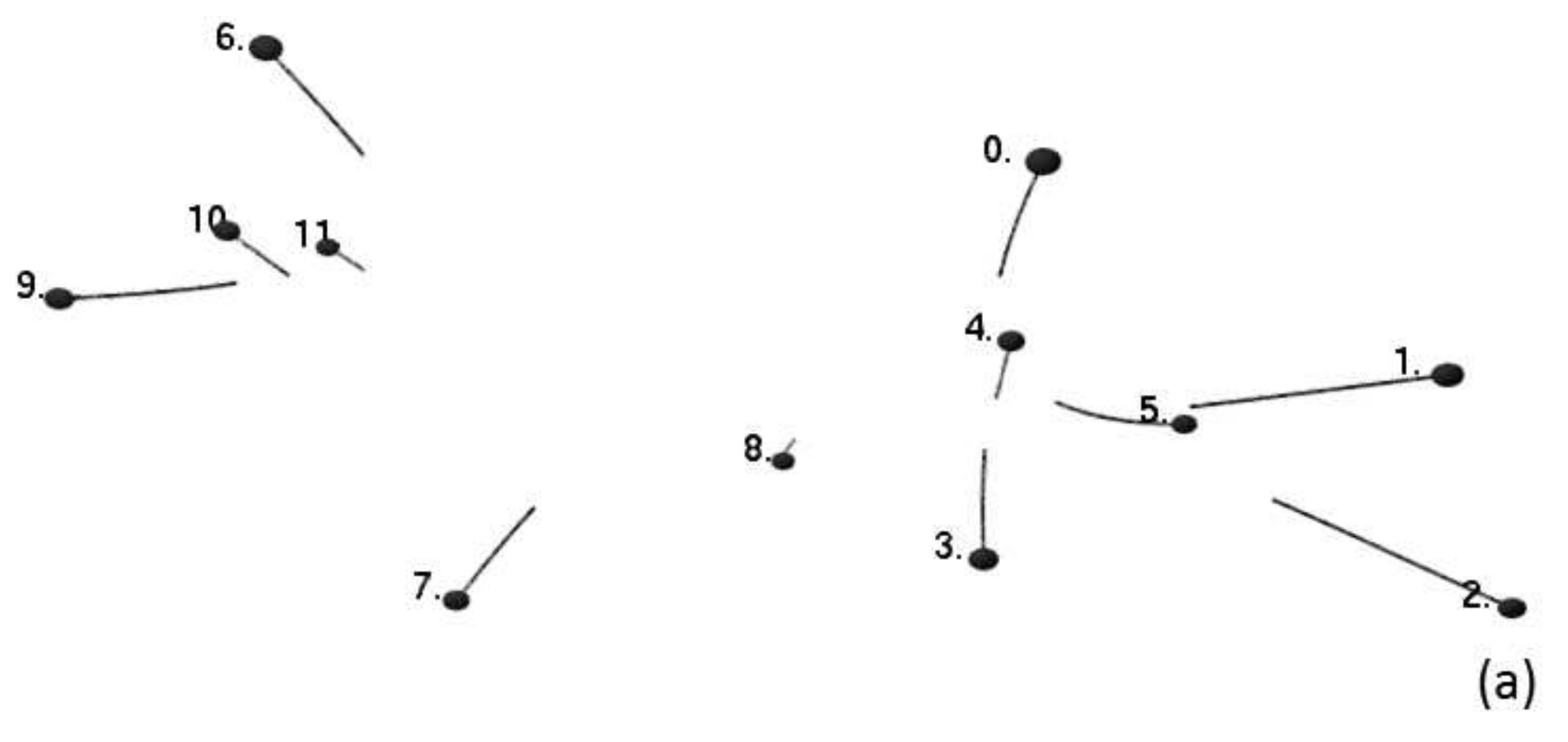}\hfill \includegraphics[width=6.5cm]{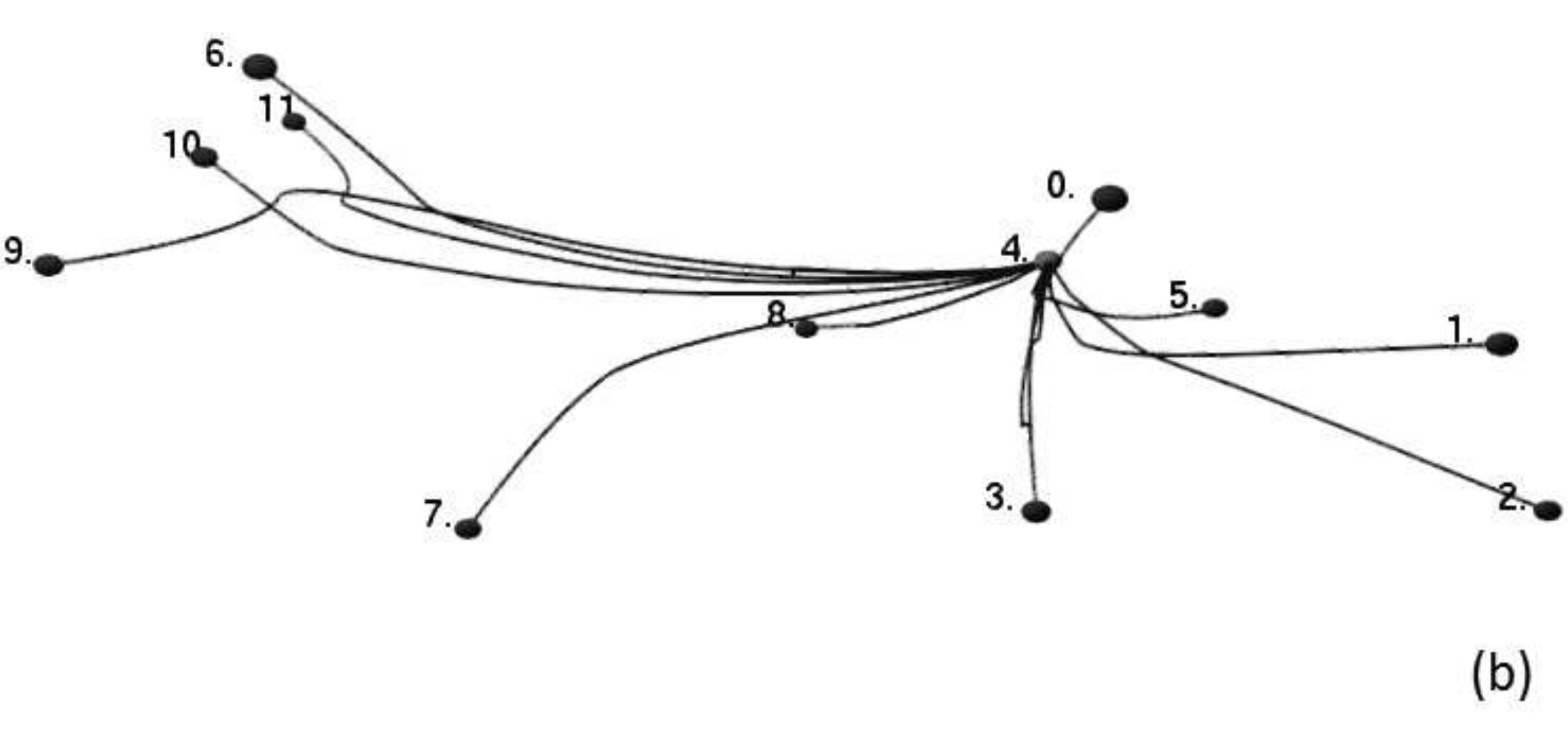}
\caption{Axonal network made in the first numerical experiment with $j_{ext}=0.1$.  (a) The initial state of network. (b) The final stage of evolution. There are $11$ connections -- $8$ positive and $3$ negative. New branches do not form new connections. It is explained in the following way. During the axons growth the situation in the axonal environment was not changed: no connections created,  the gradient of ADM concentration remains the same. For this reason influence on branch terminal and axon terminal are the same. On the scale of the initial length of branch the gradient of concentration is invariable.}\label{fig:6}
\end{figure}

\begin{figure}[ht]
\includegraphics[width=6.5cm]{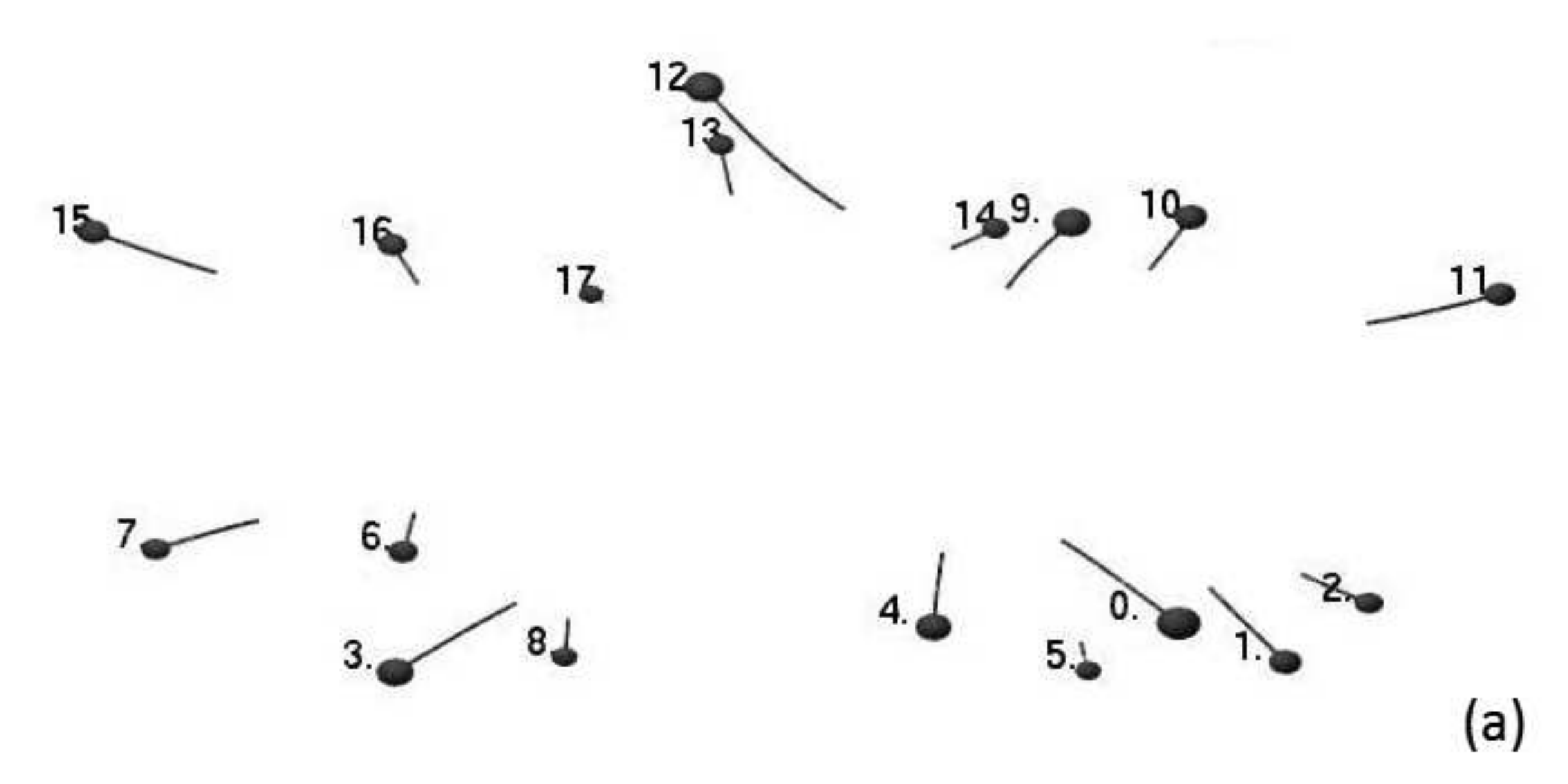}\hfill \includegraphics[width=6.5cm]{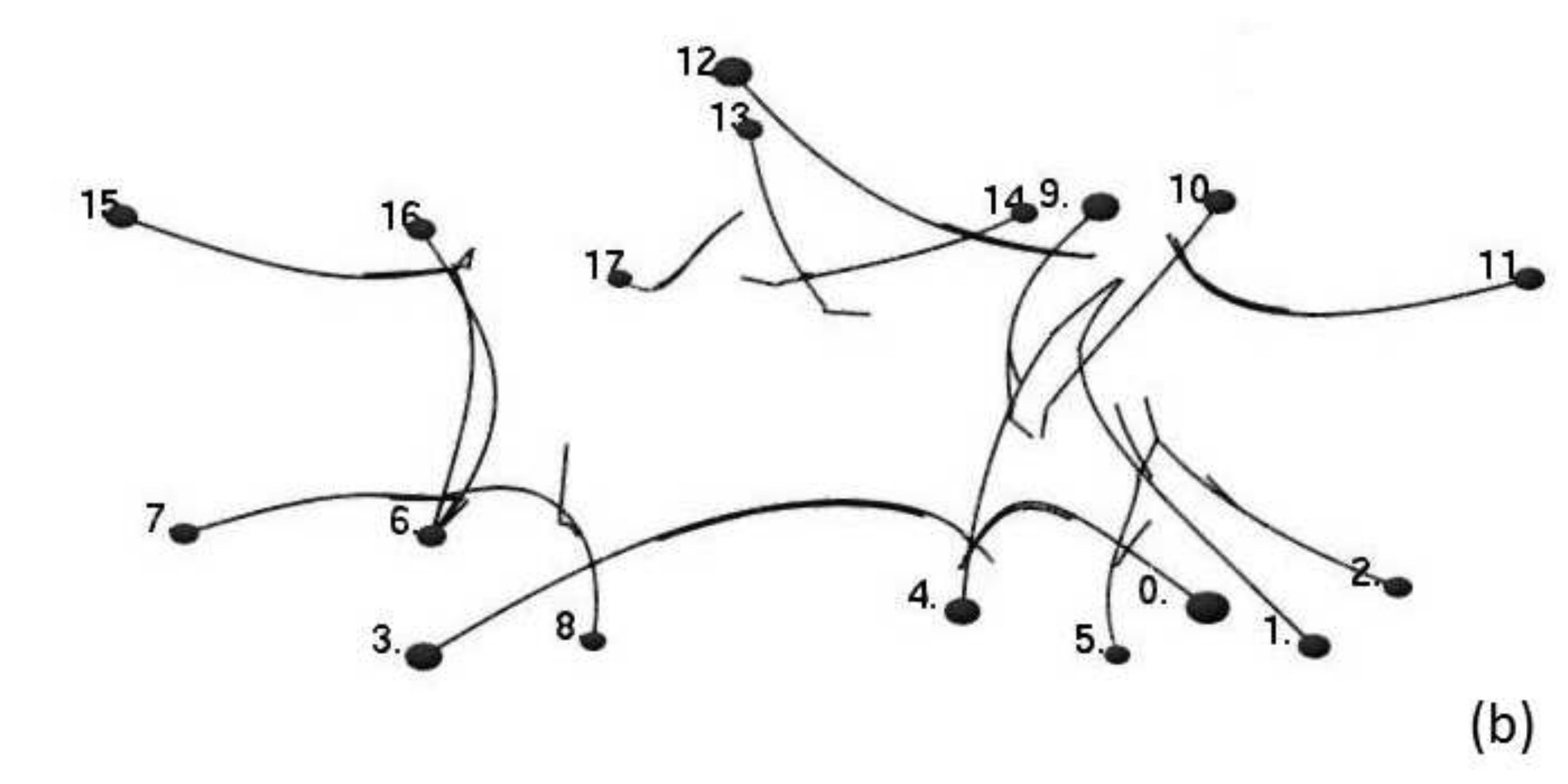}
\includegraphics[width=6.5cm]{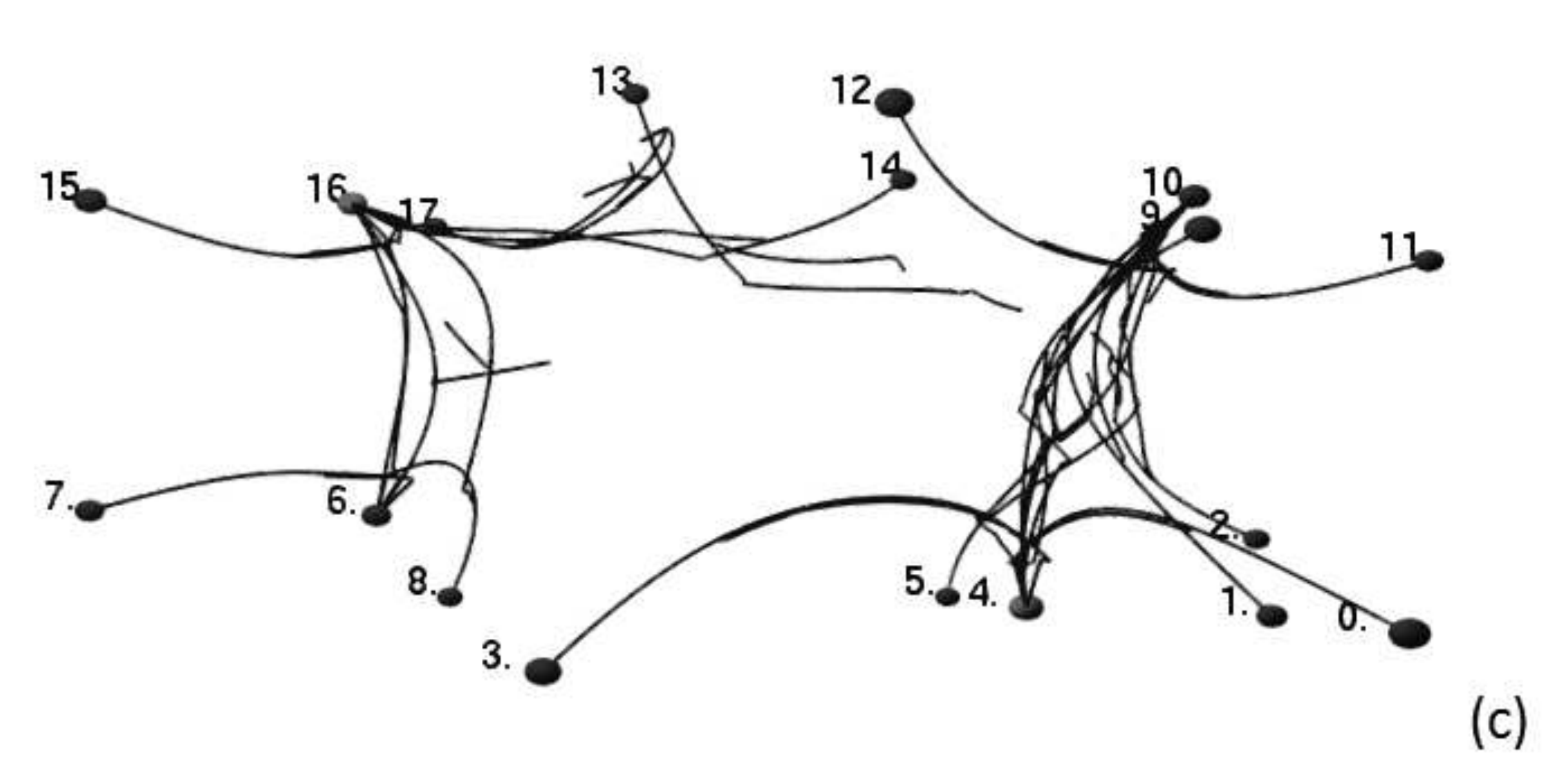}\hfill \includegraphics[width=6.5cm]{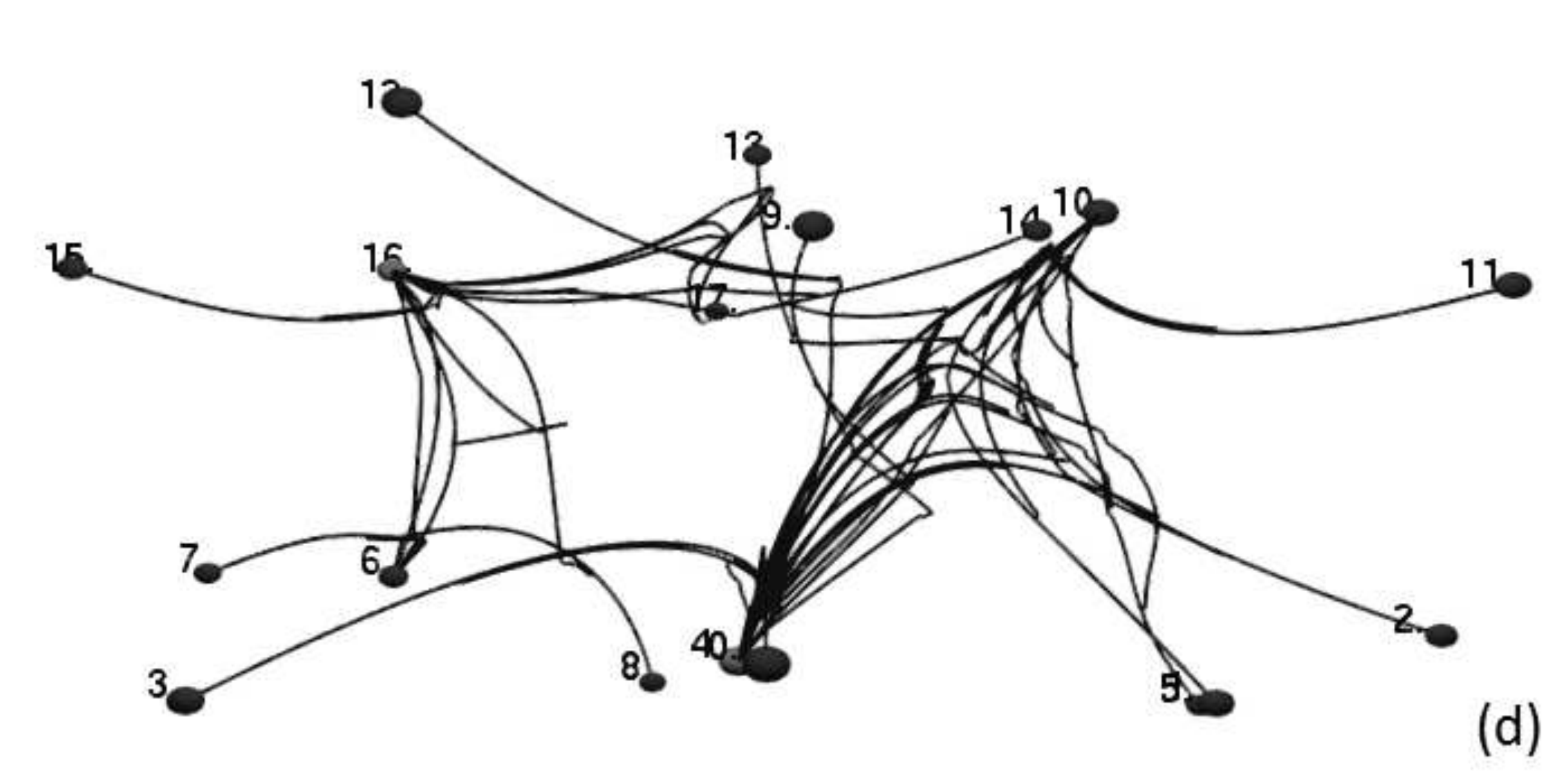}
\caption{Axonal network obtained in the second numerical experiment with $j_{ext}=0.5$.  (a) Initial state of the network. (b), (c) Intermediate stages. (d) Final stage. There are $32$ connections -- $15$ positive and $17$ negative. $10$ axons made connections without branching, $4$ axons made connections with help of single branch, $2$ axons -- with two branches, and $2$ axons made connections with help of $3$ additional branches.}\label{fig:7}
\end{figure}

We calculated amount of branching points and number of formed branches. We also obtained concentrations of AGM at branching points, neuron's activity at moment of branching, velocity of axon's terminal, length of branch and numbers of neurons between which connections are formed and type of connection. The data obtained in $9$ numerical experiments are arithmetically averaged and represented as plots. Except numerical data we show results of numerical analysis axonal network with neurons and axons to see the network obtained (see Fig. \ref{fig:6} and \ref{fig:7}).

\begin{figure}[ht]\vspace*{-14em}
\includegraphics[width=5.7cm]{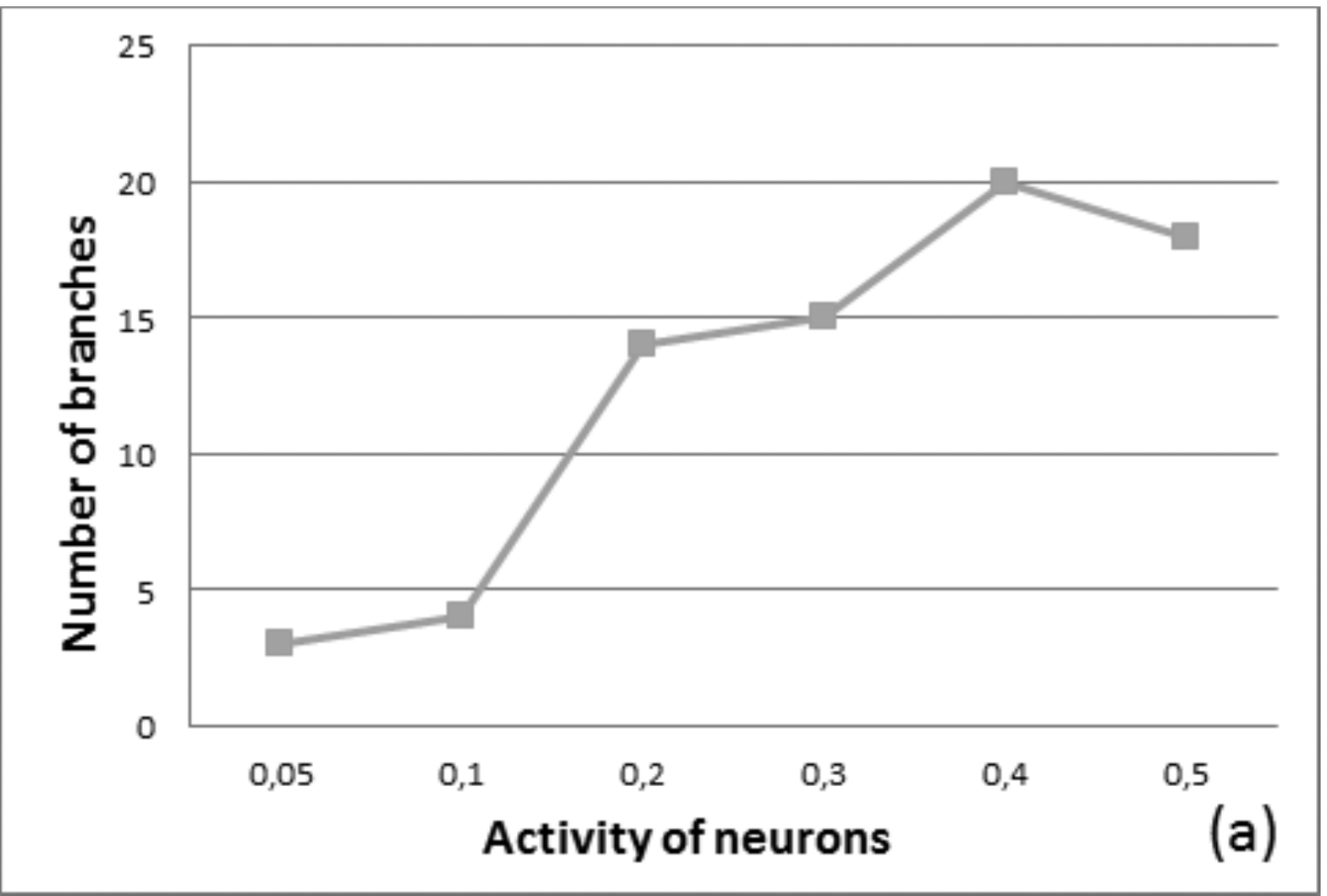}%
\hspace*{-2.6em}\includegraphics[width=5.7cm]{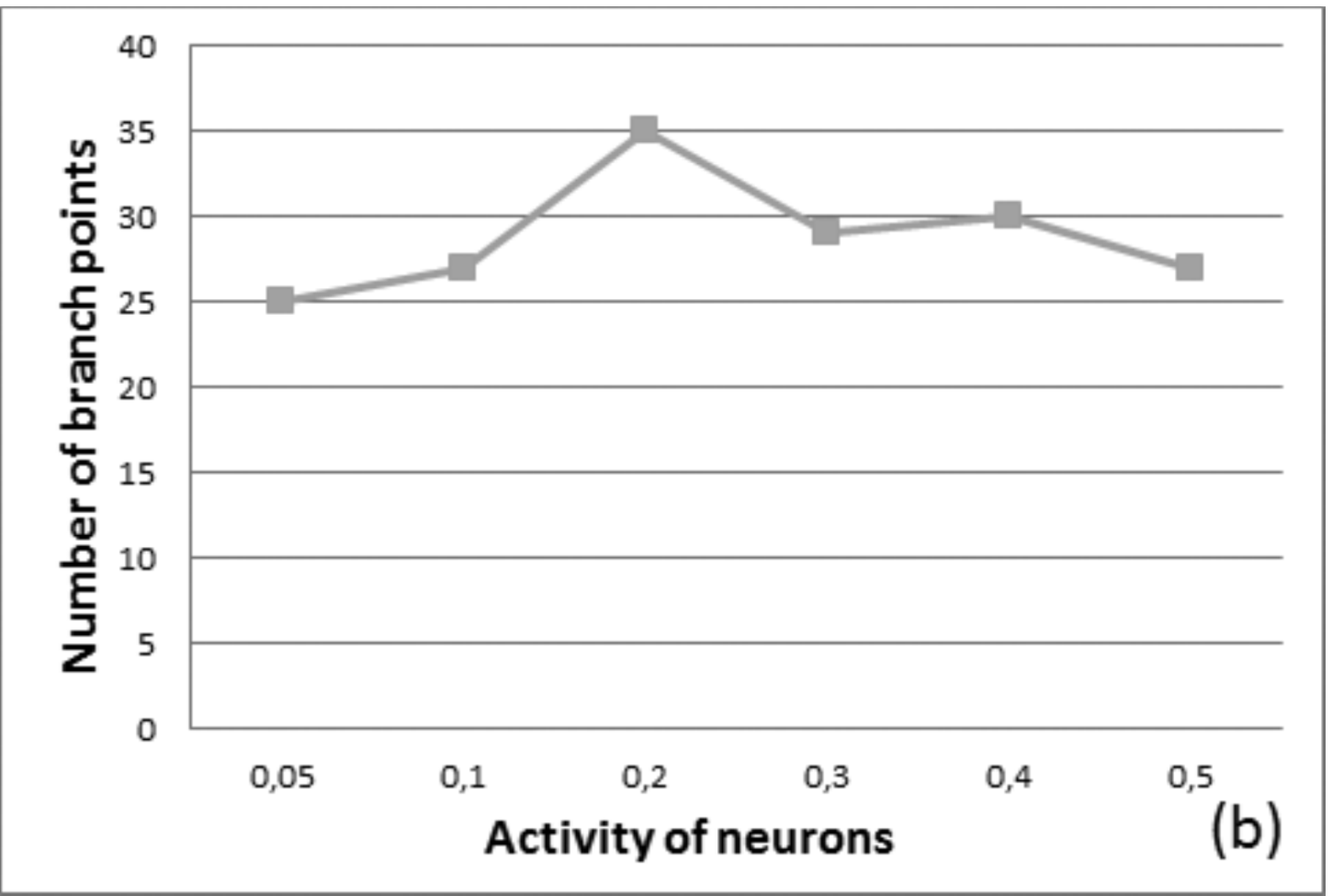}%
\hspace*{-2.6em}\includegraphics[width=5.7cm]{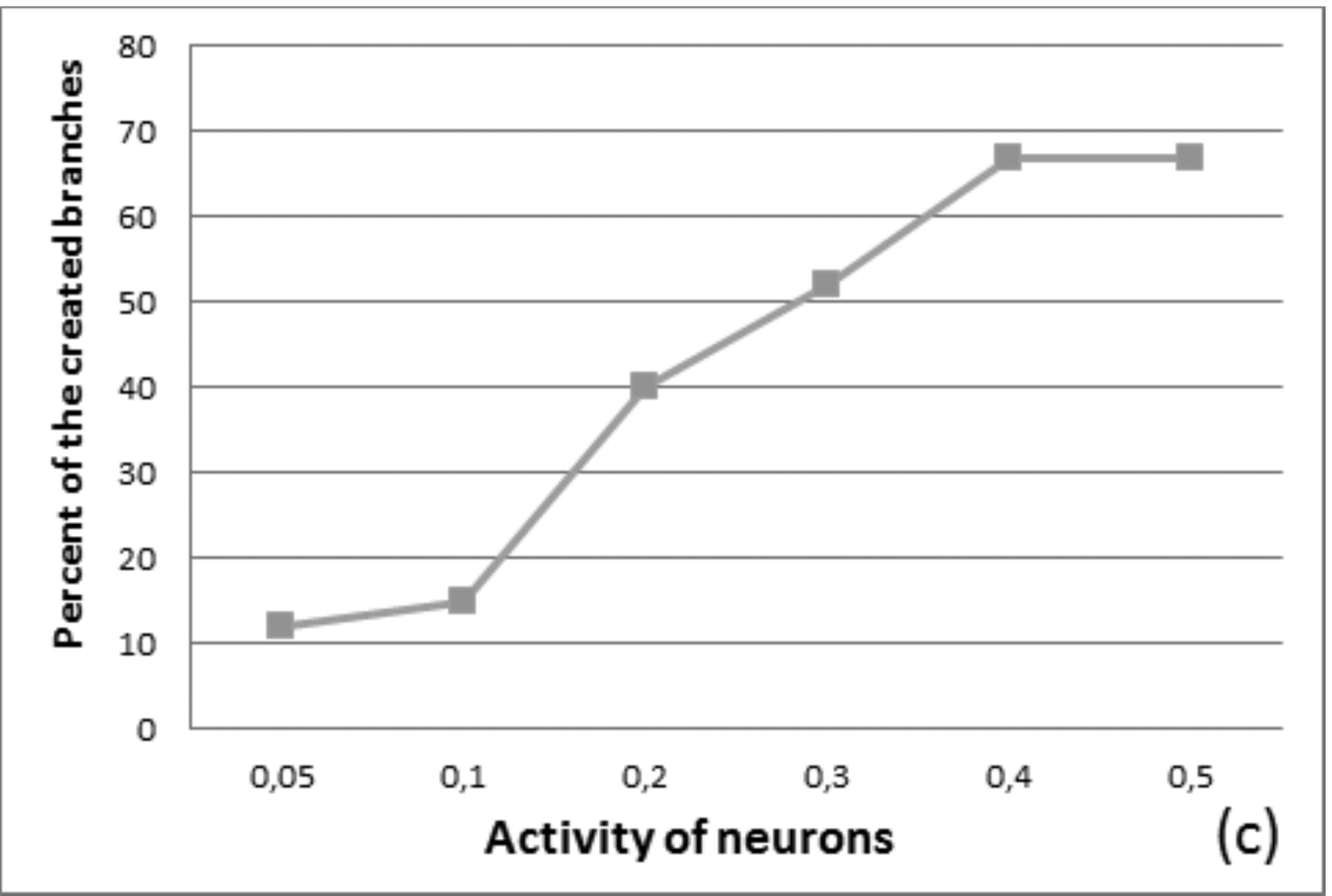}
\caption{The results of the first experiment. The activity dependence on the number of created branches (a), the number of the points of branches (b), and percent of created branches (c). }\label{fig:3}
\end{figure}
\begin{figure}[ht]\vspace*{-16em}
\includegraphics[width=5.7cm]{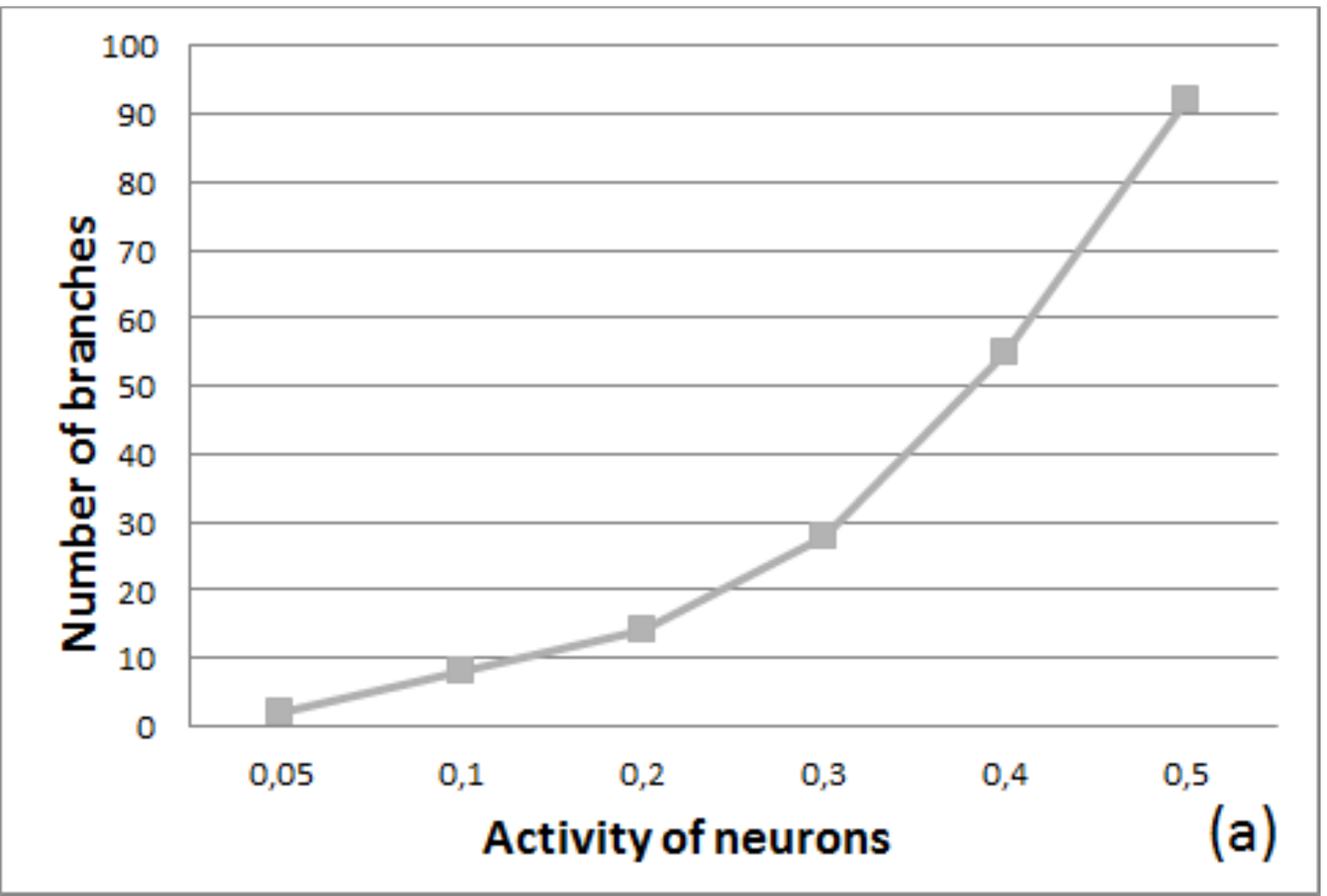}%
\hspace*{-2.6em}\includegraphics[width=5.7cm]{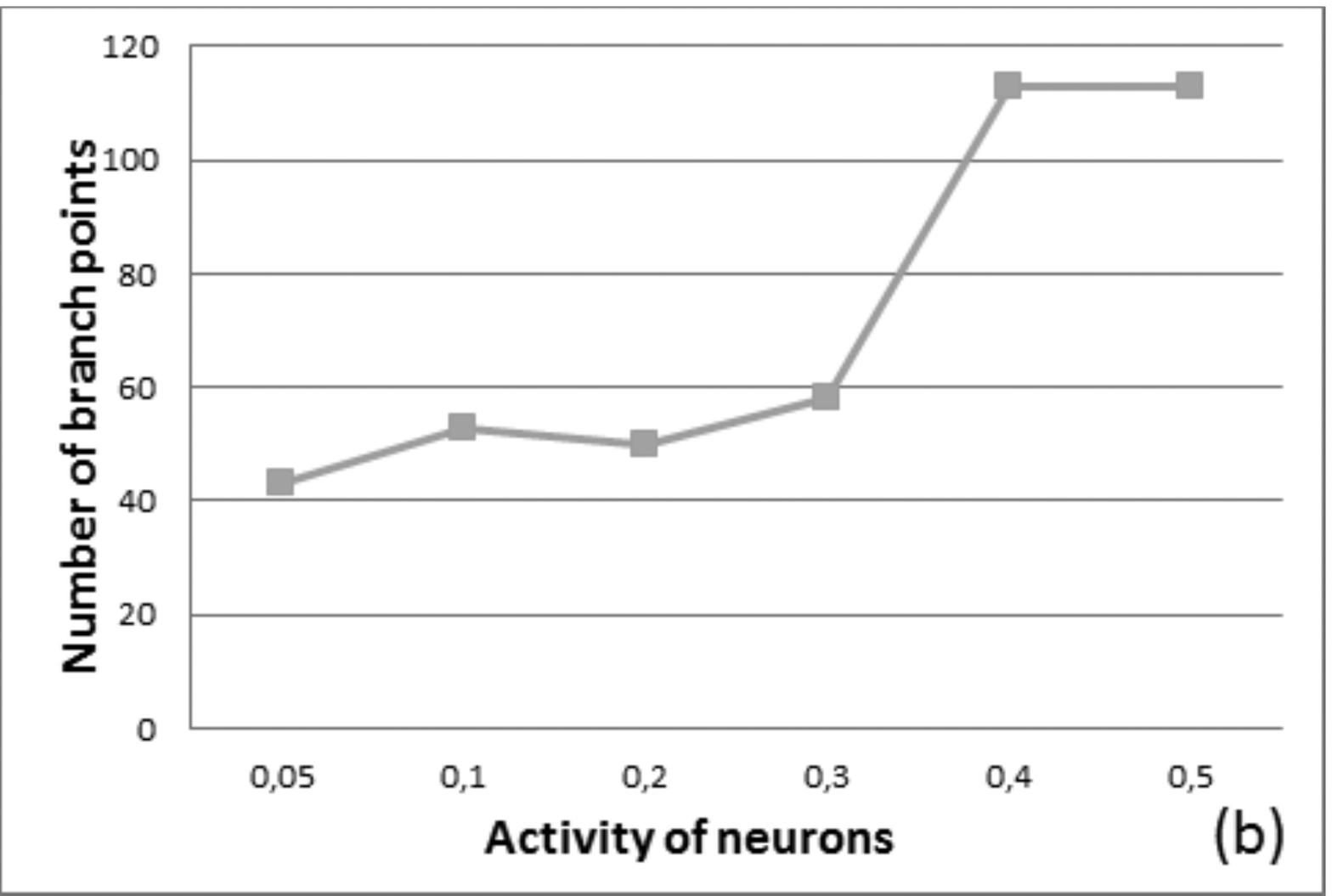}%
\hspace*{-2.6em}\includegraphics[width=5.7cm]{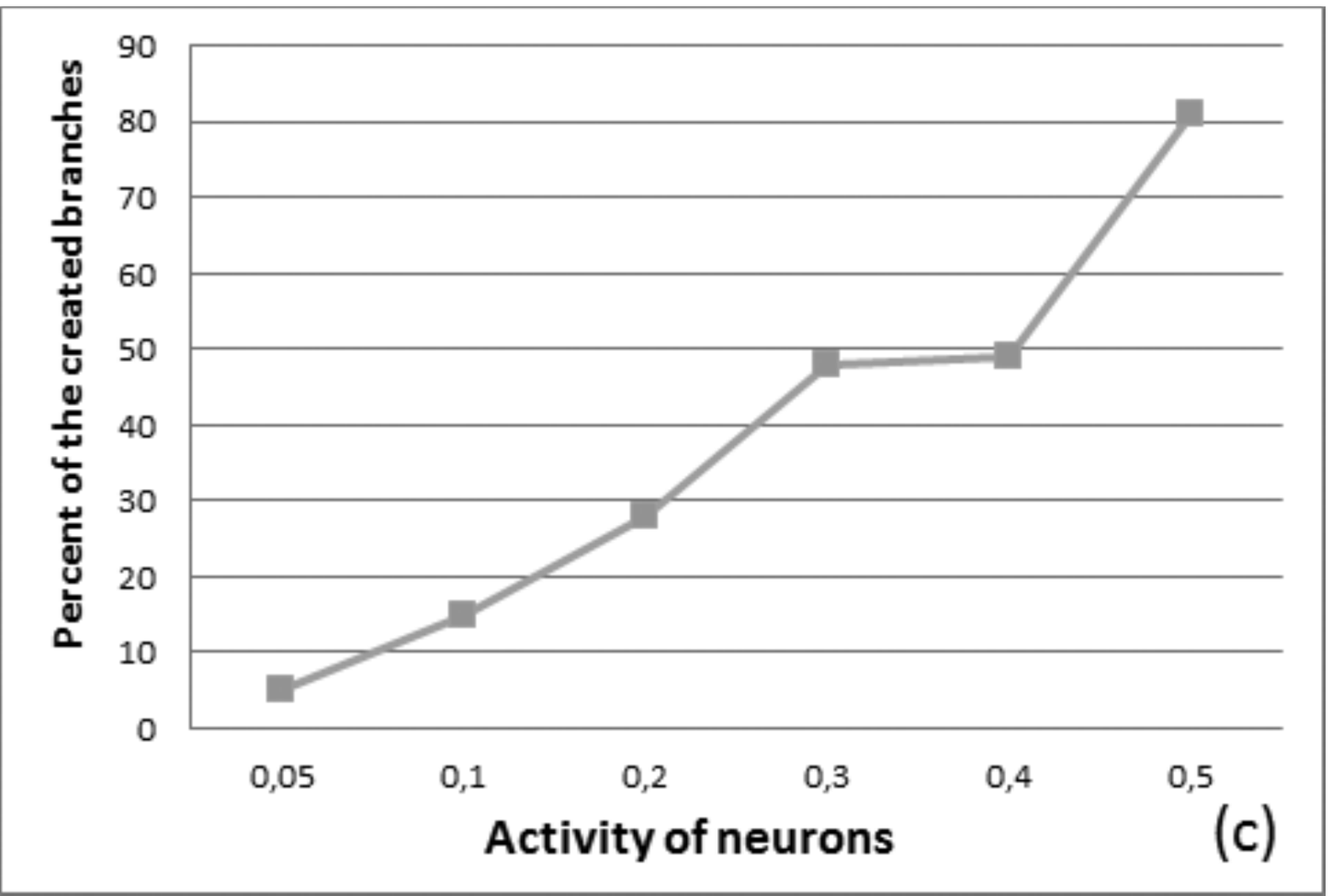}
\caption{The results of the second experiment. The activity dependence on the number of created branches (a), the number of the points of branches (b), and percent of created branches (c).}\label{fig:4}
\end{figure}
\begin{figure}[ht]\vspace*{-16em}
\includegraphics[width=5.7cm]{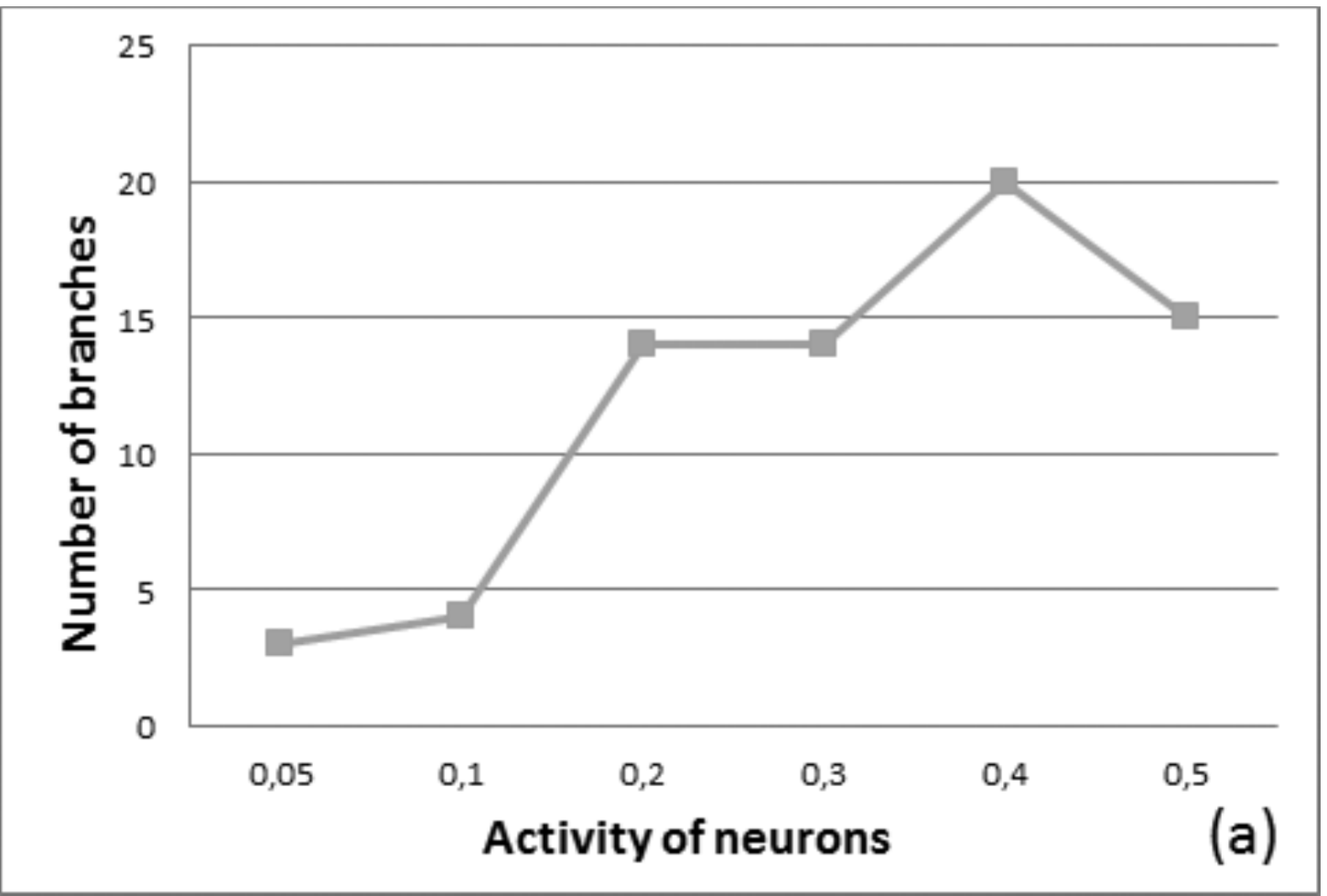}%
\hspace*{-2.6em}\includegraphics[width=5.7cm]{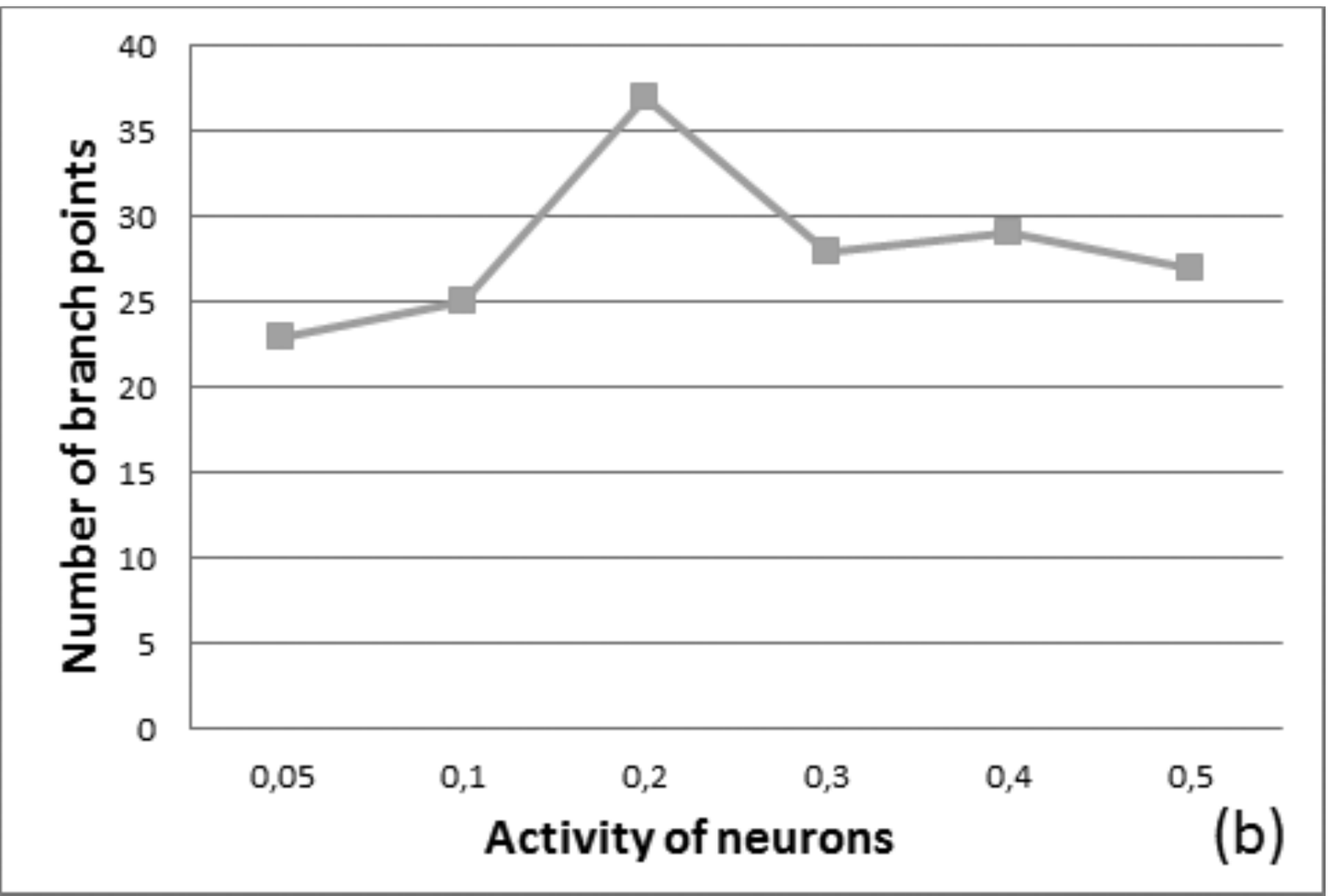}%
\hspace*{-2.6em}\includegraphics[width=5.7cm]{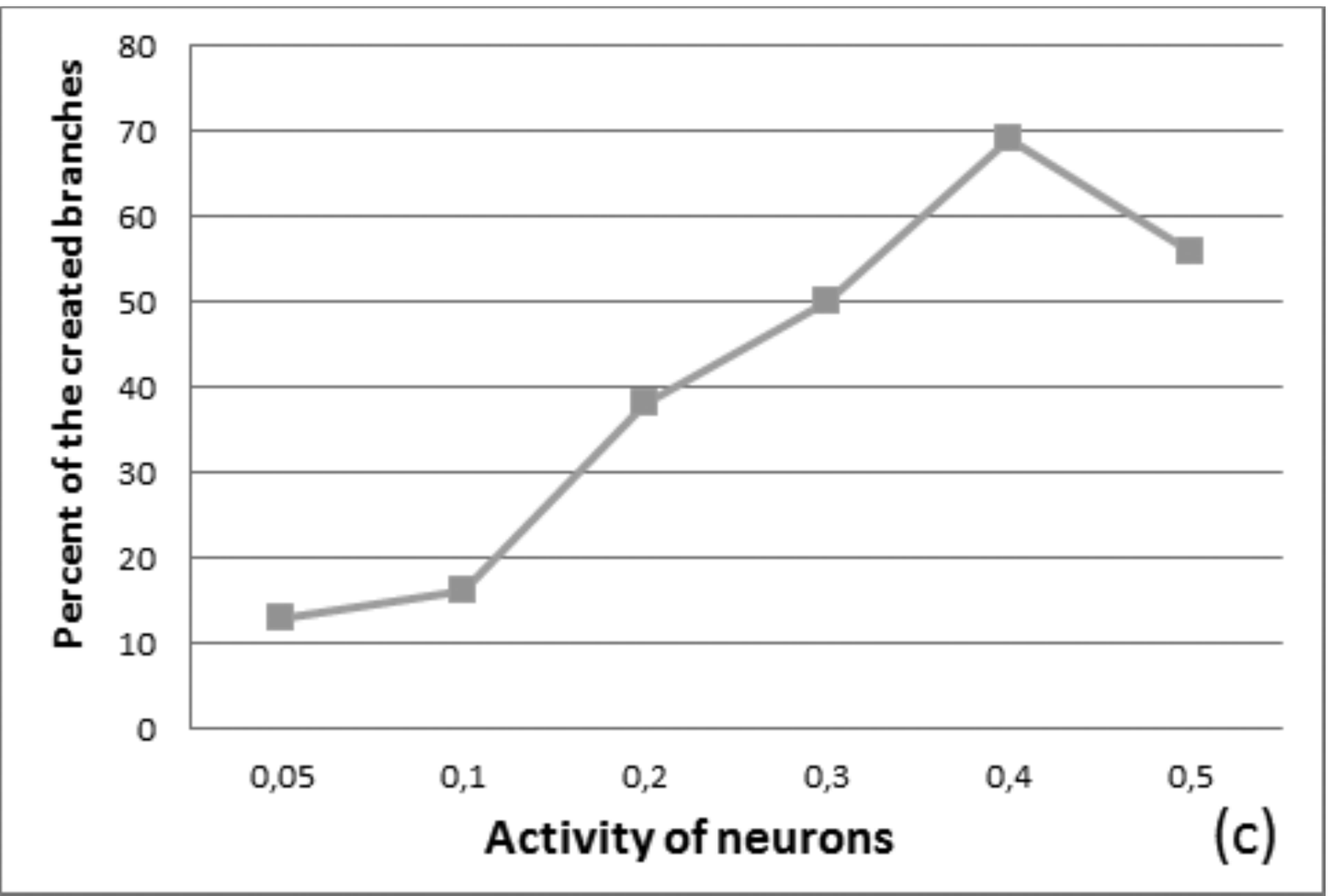}
\caption{The results of the third experiment. The activity dependence on the number of created branches (a), the number of the points of branches (b), and percent of created branches (c).}\label{fig:5}
\end{figure}

In the first numerical experiments with activity $0.1$ and $0.5$ number of branching points is $27$. This is lowest value except the case with $j_{ext}=0.05$. For activity $j_{ext}=0.1$ the number of the branches is smaller and for this reason the number of the branches is the least. For $j_{ext}=0.5$ activity of neurons is highest. It influences velocity of axon's growth and rate of network growth because the concentration of AGM depends on the neuron's activity. For great velocity and lower time of growth the number of points is smaller. Neurons with activity  $j_{ext} = 0.2$ have the maximum number of branching points, but  when activity increases the number of branching point decreases (see Fig. \ref{fig:3}(b)). Number of generated branches becomes greater with increasing of neuron activity except the case with activity $0.5$ (see Fig. \ref{fig:3}(a)). It is also explained by linear dependence between velocity of growth and activity of neurons, the time of growth is smaller than in other cases with smaller activity.

The ratio of the number generated branches and the number of generated branch points is shown in Fig. \ref{fig:3}(c). For $j_{ext}=0.1$ only $15 \%$ from $27$ branching points create new branches.  The probability of new branch creation from the branching points increases up to $67\%$ with increasing of neuron's activity to $j_{ext} = 0.4$. A drastic increase of probability to create new branch up to $40\%$ appears for $j_{ext} = 0.2$.

In the second numerical experiment we have $18$ irregular located neurons. The experiment was made $6$ time and results were averaged.  Unlike first experiment, we used different initial length of branching $r_{b}=5\cdot 10^{-4} cm$ and another disposition of neurons.  We obtained different results. The number of created branching points  and branches becomes greater. In Fig. \ref{fig:4}(a) we show average number (over $6$ experiments) of created branches for different neuron's activity. In the first experiment we observed drastic increasing of number of created branches at  $j_{ext}=0.2$ with small variations for $j_{ext} > 0.2$, but in the second experiment we obtained dependence which looks like exponent without jump at point $0.2$.

The number of branching points (see Fig. \ref{fig:4}(b)) is distinctive comparing with first experiment. In the latter case we observed the maximum at point $0.2$ while in the second experiment -- at points $0.4,0.5$. The percent of created branches grows exponentially, but for activities $0.3$ and $0.4$ the results of first and second experiments coincides with error in $1\%$ (see Fig. \ref{fig:4}(c)).

Because different results in two experiments were obtained, we repeated first experiment but with the length of branching as in the second experiment  $r_{b}=5\cdot 10^{-4} cm$. We refer this experiment as third experiment. The results are the same as in the first experiment (see Fig. \ref{fig:5} and Fig. \ref{fig:3}). Therefore, this parameter is of no importance for network formation. We found also similar results in all experiments namely, (i) the number of created branches for  activity $j_{ext}=0.1$ equals $15\%-16\%$ and (ii) for activity $j_{ext}=0.3$ we obtained $48\%-52\%$ in all experiments.

\section{Discussion}

We developed model to describe the process of axon branching during growth of neuron network. The basis of the model is diffusion of AGM, the growth process is managed by diffusion equation. We incorporate only one type of branching -- interstitial branching from the axon's shaft.

The axon's branching is activity depended process as was shown by   \cite{Uesaka:2007:Iblsaamotab,Uesaka:2005:Adocabfamaesuosc} and \cite{Ohnami:2008:Roriacab} with help of the pharmaceutical drugs, blockades and inhibitors. In the present work we verified that this branching process is close to that obtained in real experiments by \cite{Uesaka:2007:Iblsaamotab} who applied inhibitors depressing neuron's activity. They observed growing axons and their branching process.  The inhibitor TTX (tetrodotoxin) depresses the ability of nerve fibres to conduct pulses and decreases the neuron's activity. $2.1\pm 0.7$ new brunches per one neuron appeared on 14 axons when TTX was applied. In our first numerical experiment the average number of  branches is $2.5$ for low activity $j_{ext}=0.05$ which models the inhibitor application. The work undertaken by Uesaka  et al (2007) revealed that $1.5\pm 0.4$ branches appear when APV/DNQX was applied, that corresponds to activity lower than $j_{ext}=0.05$. In case of DNQX inhibitor they obtained $1.7 \pm 0.5$ branches which corresponds to $2.25$ branches in our numerical experiment for activity $j_{ext}=0.05$. The application of APV  in culture gives $5.7\pm 1.2$ branches. It corresponds activity lying in the range from $j_{ext}=0.1$ to $j_{ext}=0.2$. The results of Uesaka et al (2007) without inhibitors correspond the neuron's activity lying in the range from $0.3$ to $0.4$. The second numerical experiment gives the same results as the first experiment when TTX,   DNQX or APV/DNQX were applied.  Without inhibitors we obtain neuron's activity lying in the range from $0.1$ to $0.2$ which differs from the first experiment. It may be explained by the difference in the number of neurons and their dispositions. Concentration of AGM depends on number of neurons, distance between them and their activity. The greater the neurons quantity, the greater is concentration of AGM. The branching process depends on the AGM concentration around the axon's tip and the branch point.

The expected topology of the axonal network was observed in our model. In the framework of the model the AGM represents the single process that participate in the axon's branching. In reality the different types of molecules starts different branching mechanisms, influence different parts of axon, are observed in different parts of nervous system and form specific forms of branching. In our model we do not take into account the specific properties of each type of molecules. The branching process may be accompanied by contraction of axon branches during growth of axonal  network and competition of branches at projection to the target axon. In these cases there is connection of single branch  axon with target neuron, the other branches are contracted or change their growth trajectory.

Realization of these phenomena is not so difficult but the main problem for analysis is time of calculations which becomes very huge even for system made of dozen neurons. In future we intend to make calculations for system with hundreds axons and take into account different phenomena in the branching process (competitive between axonal branches, pruning branches).

%\bibliography{/home/nail/Library/neuro}
%\bibliographystyle{ws-jin}
%% \bibliographystyle{apsrev}
%%\bibliographystyle{cell}
%\bibliographystyle{chicago}
%%\bibliographystyle{harvard}
%%\bibliographystyle{gost71u}
%%\bibliographystyle{gost705}
\end{document}